\documentclass[11pt]{article}
\usepackage[a4paper,margin=1.1in]{geometry}
\usepackage{libertinus}
\usepackage[T1]{fontenc}
\usepackage{amsmath,amssymb,amsthm,mathtools}
\usepackage{enumitem}
\usepackage[hidelinks]{hyperref}

\theoremstyle{plain}
\newtheorem{theorem}{Theorem}
\newtheorem{corollary}{Corollary}
\theoremstyle{definition}
\newtheorem{problem}{Problem}
\newtheorem{lemma}{Lemma}
\newtheorem{definition}{Definition}
\newtheorem{fact}{Fact}
\newtheorem{proposition}{Proposition}

\newcommand{\cP}{\mathcal P}
\newcommand{\E}{\mathbb E}
\newcommand{\cG}{\mathcal G}
\newcommand{\cW}{\mathcal W}
\newcommand{\cT}{\mathcal T}
\newcommand{\cA}{\mathcal{A}}
\newcommand{\cC}{\mathcal{C}}
\newcommand{\cE}{\mathcal E}
\newcommand{\GG}{\mathbb G}
\newcommand{\Real}{\mathbb R}
\newcommand{\Sym}{\mathrm{Sym}}
\newcommand{\dist}{\operatorname{dist}}
\newcommand{\rej}{\operatorname{rej}}
\newcommand{\cost}{\operatorname{cost}}
\newcommand{\clip}{\operatorname{clip}}
\newcommand{\ones}{\mathbf 1}
\newcommand{\Mt}{\widetilde M}
\newcommand{\tup}[2]{#1^{#2}_{\neq}}
\newcommand{\ip}[2]{\langle #1,#2\rangle}
\newcommand{\tn}[1]{\lVert #1\rVert_{\cT}}
\newcommand{\Val}{\operatorname{Val}}
\newcommand{\OPT}{\operatorname{OPT}}

\title{Testing Implies Distance Estimation with Polynomial Loss}

\author{Gaia Carenini\footnote{Trinity College Cambridge, Department of Pure Mathematics and Mathematical Statistics, Centre for Mathematical Sciences, Wilberforce Road, Cambridge CB3 0WA, United Kingdom. Email: \texttt{gc645@cam.ac.uk}} \quad \&  \quad Cameron Seth \footnote{Cheriton School of Computer Science, University of Waterloo, 200 University Ave W, Waterloo, ON N2L 5Z5, Canada. Email: \texttt{cjmpseth@uwaterloo.ca}}}
\date{\today}
\begin{document}
\maketitle
\begin{abstract}
A tester decides whether a graph has a property or is \emph{far} from having it. 
A distance estimator asks for more: it must approximately determine how
many edges need to be changed.
A property is said to be \emph{testable} if there is a tester with a number of queries bounded independently of the graph's order for each conceivable distance. Similarly, if a property admits an estimator with a number of queries bounded independently of the graph's order, it is said to be \emph{estimable}.

Fischer and Newman proved that, in the
dense graph model, every testable property is also estimable. We show that the
passage from testing to estimation requires only a polynomial increase in
query complexity. This resolves a long-standing open problem in property testing, previously tackled by Fiat and Ron, by Hoppen, Kohayakawa, Lang, Lefmann, and Stagni, and by Gishboliner, Kushnir, and Shapira. 
\end{abstract}

\section{Introduction}
\label{sec:introduction}

\subsection{Property testing in the dense graph model}

Deciding whether a large object satisfies a given property can be very expensive,
both in computation and in the amount of information that must be read.
Property testing studies a relaxation of this task: distinguishing objects
that satisfy the property from those that require a substantial proportion
of their entries to be changed before they satisfy it. The aim is to make
this distinction by inspecting only a (very) small part of the input. We refer
the reader to Goldreich's book~\cite{Goldreich} for a general introduction to the area.

We work in the \emph{dense graph model}, also called the adjacency matrix
model. A graph on a vertex set $V$ is a function
$G:\binom V2\to\{0,1\}$, and we write $\cG_V$ for the set of such graphs.
For $n=|V|\ge2$, the normalized edit distance between $G,H\in\cG_V$ is
\[
    d(G,H)
    =\frac{|E(G)\triangle E(H)|}{\binom n2}.
\]
A \emph{graph property} $\cP$ is a collection of graphs closed under
isomorphism. Let $\cP_V=\cP\cap\cG_V$; we define
\[
    \dist(G,\cP)=\min_{H\in\cP_V}d(G,H)
\]
when $\cP_V\ne\emptyset$, and set $\dist(G,\cP)=1$ otherwise.
Thus, $\dist(G,\cP)$ is the smallest fraction of pairs whose adjacency
status must be changed to obtain a graph in $\cP$ on the same vertex set.
We say that $G$ is \emph{$\varepsilon$-far}  from $\cP$ if
$\dist(G,\cP)\ge\varepsilon$. Graphs of size less than two can be
handled separately.

An algorithm receives the order $n$ of the input graph and may query
whether a given pair of vertices is an edge. An \emph{$\varepsilon$-tester}
for $\cP$ accepts every graph in $\cP$ with probability at least $2/3$ and
rejects every graph that is $\varepsilon$-far from $\cP$ with probability
at least $2/3$. No requirement is imposed on graphs whose distance lies
strictly between $0$ and $\varepsilon$. The property is \emph{testable}
if, for every $\varepsilon>0$, such a tester exists with a number of
queries bounded independently of $n$.

Throughout the paper, $q(\varepsilon)$ denotes an upper bound on the
number of \emph{edge queries}, valid for every input size. We often use
\emph{canonical testers}, which sample a uniformly random set of vertices
and inspect its induced subgraph. The canonization theorem of Goldreich
and Trevisan~\cite{GT} converts a tester making $q$ edge queries into a
canonical tester sampling $O(q)$ vertices. Conversely, inspecting the
subgraph induced by $s$ vertices uses at most $\binom s2$ edge queries.
These polynomial conversions allow us to pass between the two models
without changing whether a bound is polynomial. We nevertheless keep
track of the distinction, using $s$ for the size of a canonical sample.

Goldreich, Goldwasser and Ron~\cite{GGR} introduced graph property testing in
this model, and proved that a broad class of partition properties,
including $k$-colourability and properties prescribing a large cut, are
testable. Subsequent work established general criteria for testability.
In particular, Alon and Shapira~\cite{AS} proved that every
\emph{hereditary} graph property, meaning one closed under taking induced
subgraphs, is testable.

Regularity methods play a central role in these general results. They
replace a large graph with a bounded description that preserves enough information to detect violations of the property. The quantitative
cost of employing these methods  can, however, be very large: Szemer\'edi's regularity lemma has
tower-type bounds, and Gowers~\cite{Gowers} showed that such dependence
is unavoidable for the lemma itself. This leaves open the possibility
that a particular testing problem admits a much better bound through
a more direct argument or a weaker approximation theorem. For example,
the regularity-based argument of R\"odl and Duke~\cite{RD} for
$k$-colourability was followed by polynomial testing bounds obtained
without the full regularity lemma~\cite{GGR, BS, KS}.

Understanding when such improvements are possible is a central question
in graph property testing. Related work of Carenini, Seth and
Yoshida~\cite{CSY} gives a quantitative container characterization of
one-sided testability, connecting prescribed testing complexity to
container-like descriptions. This develops a different route to quantitative
testing results, building on the hypergraph container method
of Balogh, Morris and Samotij~\cite{BMS} and Saxton and
Thomason~\cite{ST}. Here, we study the quantitative relationship between
testing and the stronger task of distance estimation.

\subsection{Distance estimation and previous bounds}

A tester distinguishes distance zero from distance at least
$\varepsilon$. A distance estimator must approximate the distance on
every input. More precisely, an \emph{$\varepsilon$-estimator} for $\cP$
outputs a number $\widehat d$ satisfying
\[
    \Pr\bigl[|\widehat d-\dist(G,\cP)|\le\varepsilon\bigr]
    \ge\frac23
\]
for every graph $G$. We call $\cP$ \emph{estimable} if it admits such an
algorithm with query complexity depending only on $\varepsilon$.
Estimability implies testability, after changing the accuracy by a
constant factor.

Parnas, Ron and Rubinfeld~\cite{PRR} initiated the systematic study of
tolerant testing and distance approximation. Testing need not imply
distance estimation in other models~\cite{FF}. In the dense graph
model, however, Fischer and Newman~\cite{FN} proved that every testable
property is estimable. Their proof uses regularity and incurs a
tower-type loss in converting a tester into an estimator. The natural
quantitative question is how much of this loss is necessary.

This question was first studied for hereditary properties by Hoppen,
Kohayakawa, Lang, Lefmann and Stagni~\cite{HKLFS} and by Fiat and
Ron~\cite{FR}. In particular, can a hereditary property that is testable
with polynomially many queries also be estimated with polynomially many
queries? The following formulation asks for a quantitative conversion
for arbitrary testing bounds.

\begin{problem}[Testing versus estimation]
\label{prob:testing-estimation}
Suppose that a graph property $\cP$ has an $\varepsilon$-tester using
$q(\varepsilon)$ queries. Can its distance be estimated to additive
accuracy $\varepsilon$ using
$\operatorname{poly}\bigl(\max\{1,q(c\varepsilon)\},1/\varepsilon\bigr)$ queries, for an absolute constant $c>0$?
\end{problem}

For hereditary properties, earlier bounds are most precisely expressed
using induced-removal lemma parameters. At proximity $\eta$, these are numbers
$M=M(\eta)$, $\delta=\delta(\eta)>0$ and $n_0=n_0(\eta)$ such that
every $\eta$-far graph $G$ of order at least $n_0$ contains some
$F\notin\cP$, with $|V(F)|\le M$, for which a uniformly random injective
map from $V(F)$ to $V(G)$ induces a copy of $F$ with probability at least
$\delta$. Such parameters give a one-sided tester with vertex sample
size $O(\max\{n_0,M/\delta\})$.

Hoppen, Kohayakawa, Lang, Lefmann
and Stagni~\cite{HKLFS} obtained distance estimators with a doubly
exponential bound in terms of these removal parameters. Gishboliner,
Kushnir and Shapira~\cite{GKS} improved this to a singly exponential bound.

\begin{theorem}[Gishboliner--Kushnir--Shapira~\cite{GKS}]
\label{thm:gks-hereditary}
Let $\cP$ be hereditary. For an absolute constant $c_0>0$, let
$M,\delta,n_0$ be induced-removal parameters at proximity
$c_0\varepsilon$. Then, $\cP$ has an $\varepsilon$-estimator with query
complexity
\(2^{\operatorname{poly}(M/\delta,\log n_0)}\).
\end{theorem}

Using the same tools,
Gishboliner, Kushnir and Shapira also obtained a doubly exponential
conversion for general graph properties~\cite{GKS}. Thus, the remaining
question was whether a polynomial loss suffices.

\subsection{Our main result}

We prove that every tester in the dense graph model can be converted
into a distance estimator with polynomial loss.

\begin{theorem}[Testing implies estimation with polynomial loss]
\label{thm:main}
There is an absolute constant $c>0$ with the following property.
If $\cP$ has an $\varepsilon$-tester with edge-query complexity
$q(\varepsilon)$ for every $\varepsilon\in(0,1)$, then it has an
$\varepsilon$-estimator with edge-query complexity
$\operatorname{poly}\bigl(\max\{1,q(c\varepsilon)\},1/\varepsilon\bigr)$.
\end{theorem}

In particular, a graph property is testable with
$\operatorname{poly}(1/\varepsilon)$ queries if and only if its distance
can be estimated with $\operatorname{poly}(1/\varepsilon)$ queries.
The theorem applies to general graph properties and allows the
original tester to be adaptive and to have two-sided error.

For natural properties, the explicit dependence on $1/\varepsilon$
can be absorbed into the testing complexity. Following~\cite{GKS}, call
$\cP$ \emph{unnatural} if there exists $\varepsilon_0>0$ such that, for
every $0<\eta<\varepsilon_0$ and all sufficiently large $n$, either
$\cP_{[n]}=\emptyset$ or every $n$-vertex graph has distance at most
$\eta$ from $\cP$. A property is \emph{natural} if it is not unnatural.
A hereditary property is \emph{non-trivial} if it contains graphs of
arbitrarily large order and omits at least one graph.

\begin{corollary}\label{cor:main}
If $\cP$ is natural, or is a non-trivial hereditary property, then the
bound in Theorem~\ref{thm:main} simplifies to
\(\operatorname{poly}\bigl(q(c\varepsilon)\bigr)
\)
for all sufficiently small $\varepsilon$.
\end{corollary}

Indeed, the query lower bounds in~\cite[Propositions~1.7 and~1.10]{GKS}
give $q(\varepsilon)=\Omega(1/\varepsilon)$ for the edge-query
complexity in these cases. The factor $1/\varepsilon$ in
Theorem~\ref{thm:main} can therefore be absorbed into the polynomial.

\section{Main technical contribution and comparison with previous approaches}
\label{sec:overview}

\subsection{Previous approaches: estimating through partitions}

A sufficiently regular partition of a graph, together with the edge
densities between its parts, approximately determines the densities
of small induced subgraphs. For a hereditary property, the induced
removal lemma turns this information into a way of detecting graphs
that are far from the property. This suggests estimating distance
by searching over modifications of the partition's density values.
The search asks how cheaply those values can be changed so that they
are compatible with a graph close to the property.

To justify this procedure, one must relate changes in the description
to changes in the graph itself. Suppose that the prescribed densities
change by $\gamma$ on average, weighted by the sizes of the corresponding
pairs of parts. One needs to realize the new densities with approximately
$\gamma\binom n2$ edge changes while retaining the regularity required
for counting. This \emph{continuity} between editing the description and
editing the graph is a key ingredient in the Fischer--Newman
argument~\cite{FN}. The large number of parts required by regularity
accounts for its tower-type bounds.

Hoppen et al.~\cite{HKLFS} used the weaker approximation supplied by
the Frieze--Kannan lemma~\cite{FK}, whose partitions have size
exponential in a polynomial of the inverse accuracy. Weak regularity
still controls the density of each fixed induced graph, but transferring
edits while preserving the required approximation is more delicate.
Their method combines weakly regular partitions with partitions that
have no regularity requirement, leading to a doubly exponential bound.
Gishboliner, Kushnir and Shapira~\cite{GKS} established an appropriate
continuity statement: a slightly stronger form of weak regularity before
editing is enough to retain ordinary weak regularity afterwards. This
underlies their improved bounds.

\subsection{Our approach}

Our proof begins by expressing the distance as the approximate minimum
of a single objective function built from the tester. We then show that this
minimum is attained approximately by a weighted graph described using
short labels on its vertices. Optimizing over these labels is a dense
constraint satisfaction problem, whose optimum can be estimated from
a small random sample. We overview each of these steps below. 

\paragraph{Replacing membership by a penalty.}
Fix a target accuracy $\varepsilon$ and let $\tau=\varepsilon/100$.
After amplification and canonization, the tester samples
$s=O\bigl(\max\{1,q(\tau)\}\log(1/\tau)\bigr)$ vertices and rejects according to a symmetric rule
$f:\cG_s\to[0,1]$. For an $n$-vertex graph $J$, consider the following rejection rule 
$\rej(J)=\E_{\mathbf v}f(J[\mathbf v])$,
where $\mathbf v$ is a uniformly random ordered tuple of $s$ distinct
vertices. The rule $f$ may depend on $n$; throughout the argument, $n$
and its associated rule are fixed. The tester satisfies
\[
\begin{cases}
 J\in\cP\ &\Longrightarrow\ \rej(J)\le\tau,\\
    \dist(J,\cP)\ge\tau\ &\Longrightarrow\ \rej(J)\ge\frac23.  
\end{cases}
\]

For the input graph $G$ on $V$, consider
$M(G)=\min_{J\in\cG_V}\bigl\{d(G,J)+2\rej(J)\bigr\}$.
When $\cP_V\ne\emptyset$, this minimum satisfies
\begin{equation}\label{eq:overview-penalty}
    \dist(G,\cP)-\tau
    \le M(G)
    \le\dist(G,\cP)+2\tau.
\end{equation}
For the upper bound, take a closest graph in $\cP$; its edit cost is
exactly $\dist(G,\cP)$ and its penalty is at most $2\tau$. For the lower
bound, a graph $J$ within distance $\tau$ of $\cP$ has
$d(G,J)\ge\dist(G,\cP)-\tau$, by the triangle inequality. Every other
$J$ pays a penalty of at least $4/3$, which already exceeds the maximum
possible distance. Thus the tester supplies all the information about
$\cP$ needed by this objective.

We next allow the edges of a candidate graph to have weights in $[0,1]$.
For a weighted graph $W$, let $\GG(W)$ be the random graph obtained by
rounding its edges independently, with edge probabilities given by $W$.
Extend the objective by setting
\[
    \cost_G(W)=\E\,d(G,\GG(W)),\qquad
    \rej(W)=\E\,\rej(\GG(W)),\qquad
    \cE_G(W)=\cost_G(W)+2\rej(W).
\]
Then, 
\[
    \cE_G(W)
    =\E\bigl[d(G,\GG(W))+2\rej(\GG(W))\bigr].
\]
Consequently, allowing weights does not change the minimum: every
weighted objective value is an average of ordinary graph objective
values. This exact relaxation makes it possible to approximate
candidate graphs by bounded real-valued functions.

\paragraph{Approximating a candidate using short vertex labels.}
The approximation only needs to preserve two quantities: the edit cost
and the rejection probability of the fixed tester. The first is affine
in the edge weights:
\[
    \cost_G(W)
    =\E_{uv}\bigl[G(uv)+(1-2G(uv))W(uv)\bigr].
\]
Thus, its change is measured by correlation with the single function
$\ones-2G$.

To control rejection probability, we compare two weighted graphs by
changing the $\binom s2$ edge probabilities in a sampled graph one at
a time. Each change is an edge-weight difference multiplied by a
discrete derivative of the tester's rejection rule. The important point
is how these changes can be described.

Suppose the changing edge occupies positions $1$ and $2$ in the sample.
Call positions $3,\ldots,s$ the \emph{anchors}. After rounding the other
edges, the change, which we call \emph{the derivative}, is determined by the graph $b$ on the anchors
and the two adjacency vectors $x,y$ from the endpoints to the anchors:
\[
    \partial_bf(x,y)
    =f\bigl(F_1(b,x,y)\bigr)-f\bigl(F_0(b,x,y)\bigr),
\]
where $F_t(b,x,y)$ is the graph with these prescribed adjacencies and
with the edge $12$ set to $t$. In the comparison between two weighted
graphs, a vertex may need a different adjacency vector for each of the
two endpoint roles. Recording both vectors, and symmetrizing in the
endpoints, gives a bounded test of an unordered pair.

In this way we obtain a family $\cT$ of functions on vertex pairs.
Each derivative test is specified by $O(s^2)$ bits of common data and
$O(s)$ bits at each vertex; we also include the cost test $\ones-2G$.
With $\|R\|_{\cT}=\max_{A\in\cT}\bigl|\E_{uv}R(uv)A(uv)\bigr|$,
the edge-by-edge comparison gives, for sufficiently large $n$,
\begin{equation}\label{eq:overview-counting}
    |\cE_G(W)-\cE_G(Z)|
    \le\left(1+4\binom s2\right)\|W-Z\|_{\cT}.
\end{equation}
This is the counting lemma needed by the proof.

The use of restrictions of the tester is inspired by Dwork and
Tankala~\cite[Definition~3.1 and Lemma~3.2]{DT}. Here, the tester observes
an induced graph, so its queried pairs share vertices. The endpoint
labels retain these shared incidences and make the resulting tests
compatible with sampling vertices.

A weak regularity argument now provides the required approximation.
Starting from $G$, maintain a sum of signed tests and use its clipped
value as the current approximation. Whenever a test distinguishes this
approximation from a target weighted graph $H$, add a small signed
multiple of that test to the sum. A bounded potential shows
that fewer than $\gamma^{-2}$ updates suffice to obtain
\[
    W=\clip\left(G+\gamma\sum_{j=1}^{t}B_j\right),
    \qquad t<\gamma^{-2},\qquad
    \|H-W\|_{\cT}\le\gamma,
\]
where $B_j\in\cT\cup(-\cT)$ and
$\clip(a)=\min\{1,\max\{0,a\}\}$. This is a form of weak regularity
adapted to the tester, following the energy-increment principle of
Frieze and Kannan~\cite{FK} and the boosting perspective of Trevisan,
Tulsiani and Vadhan~\cite{TTV}.

Combining the descriptions of the tests gives a label $z_v$ at each
vertex, drawn from a finite set $\Lambda$ with
$\log|\Lambda|=O(s^2\gamma^{-2})$.
The \emph{approximating weighted graph} is recovered by a fixed
symmetric function $D$:
\begin{equation}\label{eq:overview-decoder}
    W_z(uv)=D\bigl(G(uv),z_u,z_v\bigr).
\end{equation}
Thus, the labels describe the approximation relative to the original
graph $G$. Common values are included in each label, and $D$ is defined
for every pair of labels, including pairs whose common values disagree.
Every labeling therefore specifies a valid weighted graph, while the
boosting argument guarantees that some labeling approximates any
given target $H$.

Taking $\gamma$ to be a sufficiently small constant multiple of
$\varepsilon/s^2$, equation~\eqref{eq:overview-counting} gives
\[
    M(G)\le
    \Mt(G):=\min_{z\in\Lambda^V}\cE_G(W_z)
    \le M(G)+O(\varepsilon),
    \qquad
    \log|\Lambda|=O(s^6\varepsilon^{-2}).
\]
The number of possible labels may be exponential, but their length
is polynomial in $s$ and $1/\varepsilon$.

\paragraph{Estimating the optimum by sampling vertices.}
For a fixed labeling, both terms of $\cE_G(W_z)$ can be written as
averages of losses on ordered $s$-tuples of distinct vertices. The
rejection term already has this form; for the edit term, average the
edge-edit losses over the $\binom s2$ pairs in the tuple. By
equation~\eqref{eq:overview-decoder}, the loss on a tuple depends only
on its induced subgraph in $G$ and the labels of its vertices.

It follows that minimizing $\cE_G(W_z)$ is a dense $s$-ary constraint
satisfaction problem with alphabet $\Lambda$. Each vertex is a
variable, its possible labels are the alphabet, and each tuple
contributes one bounded loss. Sampling a set $U$ of vertices restricts
this same instance to $U$; the decoder and the rejection rule $f_n$
remain fixed. This is why labels on vertices are useful.

The subsampling theorem of Bernasconi, Castiglioni, Celli, Farina, and
Malavolta~\cite[Theorem~3.3]{BCCFM} approximates the optimum in expectation
using a sample whose size is polynomial in the arity, the inverse
accuracy, and the logarithm of the alphabet size. The theorem also
requires a lower bound on the original instance size, polynomial in
the same parameters. Both requirements are satisfied here. Small
input graphs can be read in full; for larger graphs, averaging a
polynomial number of independent sample optima gives the required
success probability.

The sampled optimum can be computed by exhaustive search over labels,
using only the edges inside the sampled vertex sets. Combining this
estimate with the short-description approximation and
equation~\eqref{eq:overview-penalty} yields an estimate of
$\dist(G,\cP)$ with query complexity
\[\operatorname{poly}(s,1/\varepsilon)
    =\operatorname{poly}\bigl(\max\{1,q(c\varepsilon)\},1/\varepsilon\bigr)\].
The polynomial dependence on the arity and the logarithmic dependence
on the number of labels are both essential: the arity grows with the
tester, and the label set is exponentially large. The boosting
argument establishes the existence of a suitable description, and the
estimator only computes optimum values on its samples.
\section{Preliminaries}

\label{sec:preliminaries}

We first put the tester into a form that can be applied to weighted graphs.
The weights let us describe gradual changes to a candidate repair. Independent
rounding will then show that introducing weights does not change the optimum
of the objective function introduced in Section~\ref{sec:overview}.

For every integer $k\ge0$, let $[k]=\{1,\dots,k\}$, with $[0]=\emptyset$.
For integers $a,b$, let $[a,b]=\{j\in\mathbb Z:a\le j\le b\}$; in particular,
$[3,2]=\emptyset$. Given a set $X$, write $\binom Xk$ for its $k$-element
subsets and $\tup Xk$ for its ordered $k$-tuples of pairwise distinct elements.
The set $\tup X0$ consists of the empty tuple. For $0\le k\le n$, write
$(n)_k=n(n-1)\cdots(n-k+1)$, with $(n)_0=1$.
For distinct $u,v\in X$, write $uv=\{u,v\}$. All logarithms have base $2$.
Unless stated otherwise, averages over a finite set are uniform.
The empty-set conventions allow a tester with $s=2$ to have no anchors.

Let $V$ be a finite set. A \emph{graph} on $V$ is a function $G:\binom V2\to\{0,1\}$. The set of these is $\cG_V$. A \emph{weighted graph} on $V$ is a function $W:\binom V2\to[0,1]$. The value $W(uv)$ represents the probability that $uv$ is an edge. The set of these is $\cW_V$. We have that $\cG_V\subseteq\cW_V$, indeed unweighted graphs are a special case of weighted graphs. Note that when $|V|<2$, $\binom{V}{2}=\emptyset$, and $\cG_V=\cW_V$ consist of the unique empty function.

For $V=[s]$, we use the following shorthands $\cG_s,\cW_s$. For $W\in\cW_V$, $\GG(W)$ is the random graph in $\cG_V$ whose edges are sampled independently at random with $\Pr[\GG(W)(e)=1]=W(e)$. If $W$ is a graph, $\GG(W)=W$. For $\mathbf v=(v_1,\dots,v_s)\in\tup Vs$, the \emph{induced weighted graph} $W[\mathbf v]\in\cW_s$ is $W[\mathbf v](ab)=W(v_av_b)$.

Let $n=|V|\geq 2$. For functions $X,Y:\binom V2\to\Real$, let $\E_e$ denote expectation over a
uniformly random unordered pair $e\in\binom V2$, and define
$\ip XY=\E_e[X(e)Y(e)]$, and $\lVert X\rVert_2=\bigl(\ip XX\bigr)^{1/2}$.
Thus, inner products measure average correlation, independently of the order
of the graph. Averages indexed by $uv$ use the same unordered-pair convention;
when ordered pairs are needed, we write them as $(u,v)$.
We use $\ones$ to denote the constant function $1$, and $\ones[\mathcal B]$
for the indicator of a condition $\mathcal B$. The \emph{distance} between graphs $G,H\in\cG_V$ is given by
\[
d(G,H)=\E_e|G(e)-H(e)|=\frac{|E(G)\triangle E(H)|}{\binom n2}\in[0,1].
\]
A \emph{graph property} $\cP$ is a set of graphs closed under isomorphism, and $\cP_V=\cP\cap\cG_V$. The \emph{distance to $\cP$} is
\[
\dist(G,\cP)=
\begin{cases}
\min_{H\in\cP_V}d(G,H)&\cP_V\ne\emptyset,\ n\ge2,\\
1&\cP_V=\emptyset,\\
0&\cP_V\ne\emptyset,\ n<2.
\end{cases}
\]
$G$ is \emph{$\varepsilon$-far} from $\cP$ if $\dist(G,\cP)\ge\varepsilon$. For graphs $G,J$ with $\cP_V\ne\emptyset$, $\dist(G,\cP)\le d(G,J)+\dist(J,\cP)$.

The next proposition replaces the internal operation of the tester by a
single rejection rule on a small induced graph. This is useful because we can
then study how that rule changes when one sampled edge changes. Amplification
makes rejection rare on graphs in the property; this will make the penalty
small for a genuine repair. The original input order $n$ remains available to
the tester, so its rule is allowed to depend on $n$.

We use the canonization theorem of Goldreich and Trevisan~\cite{GT} in the
following functional form.  

\begin{proposition}\label{prop:normal}
Let $\cP$ be a property. Fix $\tau\in(0,1/3]$ and a $\tau$-tester for $\cP$ that requires at most $q(\tau)$ queries. Then, there exist an integer $s=O(\max\{1,q(\tau)\}\log(1/\tau))\ge2$ and, for each $n\ge s$, a function $f=f_n:\cG_s\to[0,1]\cap\mathbb Q$, where $f(J)$ is the probability of rejecting when the induced graph on the $s$ sampled vertices is $J$, such that the following two conditions hold: 
\begin{enumerate}[label=(N\arabic*)]
\item $f(J\circ\sigma)=f(J)$ for all $J\in\cG_s$ and $\sigma\in\Sym([s])$, where $(J\circ\sigma)(ab)=J(\sigma(a)\sigma(b))$;
\item for every $V$ with $|V|=n$ and $J\in\cG_V$, writing $\rej(J)=\E_{\mathbf v}f(J[\mathbf v])$ with $\mathbf v$ uniform in $\tup Vs$,
\begin{equation}\label{eq:tester}
\begin{cases}
J\in\cP\ &\Rightarrow\ \rej(J)\le\tau,\\\dist(J,\cP)\ge\tau\ &\Rightarrow\ \rej(J)\ge\tfrac23 .
\end{cases}
\end{equation}
\end{enumerate}
\end{proposition}

\begin{proof}
Amplify the tester by $O(\log(1/\tau))$ independent repetitions and a majority
vote, so that its rejection probability is at most $\tau$ on members and at
least $2/3$ on $\tau$-far graphs. Apply the canonization theorem~\cite{GT} to
obtain a tester inspecting $s=O(\max\{1,q(\tau)\}\log(1/\tau))$ uniformly
sampled vertices, padding to $s\ge2$ if necessary. Average its decision rule
over all permutations of the sampled positions. This does not change the
rejection probability on any input graph, and gives (N1). The function $f_n$
is the resulting conditional rejection probability, giving (N2). With the
usual finite-randomness convention for a randomized tester, these
probabilities are rational.
\end{proof}

From now on, $n$ and its associated rule $f=f_n$ are fixed whenever we compare
graphs on $V$. In particular, restricting an optimization problem to a smaller
vertex set later in the proof will not replace $f_n$ by the tester for that
smaller order.

Recall that a function $g:\cW_V\to\Real$ is \emph{multilinear} if it is affine
in each coordinate $W(e)$ separately. Such a function treats an edge of weight
$p$ exactly as an average of an absent edge and a present edge, with weights
$1-p$ and $p$. The next fact applies this observation to all edges: a weighted
input cannot have a value smaller than every ordinary graph obtained by
rounding it. 

\begin{fact}\label{fact:multilinear}
If $g:\cW_V\to\Real$ is multilinear, then $g(W)=\E\,g(\GG(W))$ for every $W\in\cW_V$. Hence, $\min_{\cW_V}g=\min_{\cG_V}g$.
\end{fact}
\begin{proof}
We round one edge at a time. Affinity in the edge being rounded preserves the
expected value at each step, and after all steps we have an ordinary graph.
List the pairs $e_1,\dots,e_N$, $N=\binom n2$, and write $W_i=W(e_i)$. For $0\le k\le N$, let
\[g_k(W)=\sum_{x\in\{0,1\}^k}g(x,W_{k+1},\dots,W_N)\prod_{i\le k}W_i^{x_i}(1-W_i)^{1-x_i}.\]
Affinity in $W_{k+1}$ gives $g_k=g_{k+1}$, and therefore $g=g_0=g_N=\E\,g(\GG(W))$. This is a weighted average of values $g(J)$, $J\in\cG_V$, so it is at least their minimum. The minimum over $\cW_V$ exists by compactness.
\end{proof}
We now extend the tester by interpreting each edge weight as an independent
edge probability. The function $\phi$ is the rejection probability on one
weighted sample of size $s$; the function $\rej$ averages this quantity over
all samples in the full graph. Keeping these two levels separate will let us
analyze one sampled edge before averaging over the input graph. Define
$\phi:\cW_s\to[0,1]$ and $\rej:\cW_V\to[0,1]$ by
\[
\begin{cases}
\phi(Y)=\E\,f(\GG(Y))=\sum_{J\in\cG_s}f(J)\prod_{ab\in\binom{[s]}2}Y(ab)^{J(ab)}\bigl(1-Y(ab)\bigr)^{1-J(ab)},\\
\rej(W)=\E_{\mathbf v\in\tup Vs}\,\phi\bigl(W[\mathbf v]\bigr).
\end{cases}
\]

The next fact says that the extension retains relabeling symmetry and is
compatible with rounding. Thus we may either round the whole graph and run
the tester, or sample first and then round the sampled edges.

\begin{fact}\label{fact:rej}
The following properties hold:
\begin{enumerate}[label=(\alph*)]
\item $\phi(Y\circ\sigma)=\phi(Y)$ for $Y\in\cW_s$, $\sigma\in\Sym([s])$;
\item $\phi$ and $\rej$ are multilinear;
\item $\rej(W)=\E\,\rej(\GG(W))$, and on graphs this $\rej$ agrees with that of (N2).
\end{enumerate}
\end{fact}
\begin{proof}
Each assertion follows because relabeling and taking an induced sample
preserve the independent-edge model. More explicitly:
\begin{enumerate}[label=(\alph*)]
    \item $\GG(Y\circ\sigma)$ has the law of $\GG(Y)\circ\sigma$; use (N1). 
    \item Each summand of $\phi$ is a product of factors affine in distinct coordinates, and $\rej$ is an average of $\phi$ composed with coordinate projections.
    \item Fact~\ref{fact:multilinear}; on graphs $\phi(J[\mathbf v])=f(J[\mathbf v])$.
\end{enumerate}
 
\end{proof}
\section{Proofs}
\label{sec:proofs}
\subsection{Derivatives of the tester, tests, and the counting lemma}

The first step is to identify which features of a candidate repair need to
be preserved. There are only two relevant quantities: its expected edit cost
and its rejection probability. We will construct a family of pair functions
whose correlations control changes in both quantities.

Fix $0<\tau\le1/3$ and $s,f=f_n$ as in Proposition~\ref{prop:normal}.
Let $G\in\cG_V$ with $n\ge s$, and put $K=\binom s2$, the number of pairs
inspected in a canonical sample. Recall the definitions of
$\cost_G:\cW_V\to[0,1]$ and $\cE_G:\cW_V\to[0,3]$: 
\[
\cost_G(W)=\E\,d\bigl(G,\GG(W)\bigr),\qquad \cE_G(W)=\cost_G(W)+2\,\rej(W),\qquad M(G)=\min_{W\in\cW_V}\cE_G(W).
\]
The objective charges for editing the graph and for leaving a graph that the
tester is likely to reject. We first observe that the edit part is particularly
simple: on a non-edge of $G$, increasing the weight increases the cost, while
on an edge of $G$ it decreases the cost. The function $\ones-2G$ records
exactly these two signs. 
\begin{fact}\label{fact:cost}
The function $\cost_G:\cW_V\to[0,1]$ satisfies the equality   $\cost_G(W)=\ip G\ones+\ip W{\ones-2G}$. In particular, $\cost_G$ is affine and $\cost_G(J)=d(G,J)$ for $J\in\cG_V$.
\end{fact}
\begin{proof}
Compute the cost for one pair, then average over pairs and over the rounding.
For $a,b\in\{0,1\}$, $|a-b|=a+b-2ab$. So $d(G,H)=\ip G\ones+\ip H{\ones-2G}$ for graphs $H$. Take expectations, using $\E\,\GG(W)(e)=W(e)$.
\end{proof}

The next lemma explains why we can minimize this objective instead of
searching only over graphs in $\cP$. A genuine repair has a small penalty.
A graph that remains far from $\cP$ pays more than the largest possible edit
distance, while a graph already close to $\cP$ cannot save much on the cost
of a genuine repair. Independent rounding also shows that fractional edge
weights introduce no better minimum.

\begin{lemma}\label{lem:penalty}
The weighted and unweighted minima agree, i.e., 
$M(G)=\min_{J\in\cG_V}\bigl\{d(G,J)+2\rej(J)\bigr\}$.
Further, if $\cP_V\ne\emptyset$, then
$\dist(G,\cP)-\tau\le M(G)\le\dist(G,\cP)+2\tau$.
\end{lemma}
\begin{proof}
We first remove the weights using rounding, then compare an arbitrary
candidate with a closest genuine repair. By Facts~\ref{fact:cost}
and~\ref{fact:rej}, $\cE_G$ is multilinear, so
Fact~\ref{fact:multilinear} gives the equality.
For the distance bounds, assume $\cP_V\ne\emptyset$.
For the upper bound, take $H\in\cP_V$ with $d(G,H)=\dist(G,\cP)$. By \eqref{eq:tester}, $\cE_G(H)\le\dist(G,\cP)+2\tau$. For the lower bound, let $J\in\cG_V$. If $\dist(J,\cP)<\tau$, then $d(G,J)\ge\dist(G,\cP)-\dist(J,\cP)>\dist(G,\cP)-\tau$. If $\dist(J,\cP)\ge\tau$, then $2\rej(J)\ge4/3>1\ge\dist(G,\cP)$ by \eqref{eq:tester}. Either way $\cE_G(J)\ge\dist(G,\cP)-\tau$.
\end{proof}

The coefficient $2$ makes the lower-bound comparison work: a $\tau$-far graph
has penalty at least $4/3>1\ge\dist(G,\cP)$. With coefficient $1$, the
soundness guarantee would give a penalty of only $2/3$, which could be too
small to prevent such a graph from undercutting the true distance.

We next isolate the information needed to measure the effect of one sampled
edge. Designate its endpoints as positions $1$ and $2$, and call positions
$3,\dots,s$ the \emph{anchors}. Once the graph on the anchors and the two
endpoint-to-anchor adjacency vectors are fixed, the sampled graph is
completely determined except for the edge $12$. The gadget below simply
records this decomposition; it adds no further graph structure.

\begin{definition}[Gadget $F_t$]
For a graph $b\in\cG_{[3,s]}$, vectors $x,y\in\{0,1\}^{[3,s]}$ and integer $t\in\{0,1\}$, let
\(F_t(b,x,y):\tbinom{[s]}2\to\{0,1\}\)
be the graph on $[s]$ given by:
\begin{itemize}
    \item $F_t(b,x,y)(\{1,2\})=t$;
    \item $F_t(b,x,y)(\{1,a\})=x_a$ for every $a\in[3,s]$;
    \item  $F_t(b,x,y)(\{2,a\})=y_a$ for every $a\in[3,s]$;
    \item $F_t(b,x,y)(\{a,c\})=b(\{a,c\})$ for every $ a\ne c\in[3,s]$.
\end{itemize}
So $x_a$ encodes whether position $1$ is adjacent to anchor $a$, $y_a$ does the same for position $2$, $b$ is the graph induced on the anchors, and $t$ is the value of the edge $\{1,2\}$. 
\end{definition}

The derivative below measures how much the tester notices a change to the
remaining edge. It can be positive or negative, since adding an edge need not
make rejection more likely. This discrete difference is also the coefficient
of the edge probability when the other sampled edges have been fixed.
\begin{definition}[Derivative of $f$]
The \emph{derivative of $f$ in the edge $12$ at $b$} is $\partial_bf:\{0,1\}^{[3,s]}\times\{0,1\}^{[3,s]}\to[-1,1]$: 
\[
\partial_bf(x,y)=f\bigl(F_1(b,x,y)\bigr)-f\bigl(F_0(b,x,y)\bigr).
\]
\end{definition}

To compare two weighted graphs, we will replace the $K$ sampled edge
probabilities one at a time. At an intermediate stage, some positions use
weights from the first graph and others from the second. A vertex can therefore
need one adjacency vector when it occupies endpoint position $1$ and another
when it occupies position $2$. We store both vectors in a single \emph{color}.
The extra color $\bot$ marks the anchors themselves, so that a test ignores
pairs meeting an anchor.

Let
$\Omega=\{\bot\}\sqcup
\bigl(\{0,1\}^{[3,s]}\times\{0,1\}^{[3,s]}\bigr)$. For each $b\in\cG_{[3,s]}$, define the symmetric function
$\kappa_b:\Omega\times\Omega\to[-1,1]$ by 
\[
\begin{cases}
\kappa_b\bigl((x,y),(x',y')\bigr)=\tfrac12\bigl(\partial_bf(x,y')+\partial_bf(x',y)\bigr),\\\kappa_b(\bot,\omega)=\kappa_b(\omega,\bot)=0.
\end{cases}
\]
By (N1), $\partial_bf(x,y)=\partial_bf(y,x)$: exchanging positions $1$ and $2$
is a relabeling. The reason for averaging in $\kappa_b$ is different. The
role-dependent expression $\partial_bf(x,y')$ need not equal
$\partial_bf(x',y)$ when the two vertices have different vectors for their
two roles. Averaging these two orientations makes the test a well-defined
function of an unordered pair.

A \emph{coloring} $\chi:V\to\Omega$ assigns this local information to every
vertex. For each $b$ and $\chi$, define the \emph{test}
\(A_{b,\chi}(uv)=\kappa_b\bigl(\chi(u),\chi(v)\bigr).
\)
Thus a single coloring describes the response to every possible endpoint
pair, with the same anchor graph $b$. We allow all colorings; they need not
come from actual neighborhoods in one graph. This larger family contains the
random colorings used in the counting proof and is easy to encode by vertex
labels.

Set
\[
\cA=\{A_{b,\chi}:b\in\cG_{[3,s]},\ \chi:V\to\Omega\},\qquad
\cT=\cT(G)=\{\ones-2G\}\cup\cA.
\]
For $R:\binom V2\to\Real$, define
\[
\tn R=\max_{A\in\cT}|\ip RA|.
\]
This seminorm measures the largest average discrepancy visible to any of
these tests. A small value of $\tn{W-Z}$ means that the tests have little
ability to distinguish $W$ from $Z$; it does not assert that every edge weight
is close. The cost test $\ones-2G$ is included because its correlation is
exactly the change in edit cost.

Every test takes values in $[-1,1]$. A derivative test is specified by
$O(s^2)$ common bits for $b$ and $O(s)$ bits per vertex for $\chi(v)$.
The cost test is instead recovered directly from the input edge $G(uv)$.
This distinction will explain why the decoder later reads both endpoint
labels and the original edge.

\emph{Example ($s=3$, triangles).}
Consider the rejection rule $f(J)=\ones[J\text{ is a triangle}]$.
There is one anchor, so $b$ has no edges and $\partial_bf(x,y)=xy$ for
$x,y\in\{0,1\}$. Fix an anchor vertex $w$ in an ordinary graph $J$, and
write $N_J(w)$ for its neighborhood.
Give $w$ color $\bot$ and every other vertex $v$ the color
$(\ones[v\in N_J(w)],\ones[v\in N_J(w)])$. Then
\[
A_{b,\chi}(uv)=\ones[u\in N_J(w)]\ones[v\in N_J(w)].
\]
The test detects changes on pairs inside the neighborhood of $w$.
Summing these indicators over $w$ counts common neighbors of $u$ and $v$:
these are exactly the vertices that can complete the changing edge to a
triangle. The general derivative tests play the same role for an arbitrary
rejection rule.

The counting lemma says that these tests capture everything needed to
control the objective.
For some intuition towards the proof of the counting lemma, we continue with the triangle example. Let $W$ and $Z$ be two graphs that have a large difference in their rejection probabilities (and for simplicity assume the edge weights are from $\{0,1\}$). In other words, taking a random sample $S$ of $3$ vertices, the probability that the induced subgraph is a triangle is significantly different for $W$ compared to $Z.$ We wish to show that there exists $A \in \cT$ for which $\ip{W-Z}{A}$ is large.

To this end, consider a specific ordering $e_1,e_2,e_3$ of the edge locations between the $3$ sampled vertices $S,$ and consider interpolating between $W[S]$ and $Z[S]$ by starting with $W[S]$ and changing one edge at a time, in the order $e_1,e_2,e_3,$ until reaching $Z[S].$ For the sets $S$ for which the rejection rule differs for $W$ and $Z,$ there is some point in the interpolation for which the output of the rejection rule changes. By considering all such sets $S,$ there exists a step in the interpolation at which the rejection rule changes for a significant fraction of these sets.

After selecting an anchor graph $b$ on a single vertex $w,$ we construct $A_{b,\chi}$ using an interpolated version of $W$ and $Z$ to define the labeling $\chi.$ Roughly speaking, the entries in $A_{b,\chi}$ correspond to pairs of vertices $u,v$ for which the output of the rejection rule on the interpolated subgraph changes when flipping the edge $u,v.$ These entries in $A_{b,\chi}$ line up with non-zero entries in $W-Z,$ resulting in $\ip{W-Z}{A_{b,\chi}}$ being large.

More generally, $s$ is the canonical sample size and
$K=\binom s2$. There are $K$ possible edge changes in a sample; bounding each
one by the seminorm and summing gives the dependence on $K$.
\begin{lemma}[Counting]\label{lem:counting}
If $n\ge4s$ then for all $W,Z\in\cW_V$,
\[
\begin{cases}
|\rej(W)-\rej(Z)|\le2K\,\tn{W-Z},\\|\cE_G(W)-\cE_G(Z)|\le(1+4K)\,\tn{W-Z}.  
\end{cases}
\]
\end{lemma}

\begin{proof}
We interpolate between $Z$ and $W$ on one sampled pair at a time. For each
step, we express the tester's response as an average of derivative tests.
The only normalization loss comes from excluding the anchors as endpoints.

List the pairs of positions $e_1,\dots,e_K$ of $[s]$.
 For $0\le i\le K$ and $\mathbf v\in\tup Vs$ define $H_i(\mathbf v)\in\cW_s$ by $H_i(\mathbf v)(e_j)=W(v_av_b)$ if $j\le i$ and $Z(v_av_b)$ if $j>i$, where $e_j=ab$. Let $\rho_i=\E_{\mathbf v}\phi(H_i(\mathbf v))$. Then $\rho_0=\rej(Z)$ and $\rho_K=\rej(W)$.

Fix $i\ge1$, $e_i=ab$. For $Y\in\cW_s$ let $Y^{[t]}$ be $Y$ with its value on $ab$ replaced by $t\in\{0,1\}$, and $\Delta(Y)=\phi(Y^{[1]})-\phi(Y^{[0]})\in[-1,1]$. Since $\phi$ is affine in $Y(ab)$ and $H_i(\mathbf v),H_{i-1}(\mathbf v)$ differ only on $ab$,
\[
\rho_i-\rho_{i-1}=\E_{\mathbf v}\Bigl[(W-Z)(v_av_b)\ \Delta\bigl(H_i(\mathbf v)\bigr)\Bigr].
\]

By Fact~\ref{fact:rej}(a), we may relabel both the changing pair and the
sampled positions without changing the derivative value or the sampling law.
Thus we may assume $e_i=12$; each other pair $cd$ then carries a label $m_{cd}\in\{W,Z\}$ saying which function is used on it. Write $\mathbf v=(p,q,w)$ with $p=v_1$, $q=v_2$, $w=(w_c)_{c\in[3,s]}$, and let $\Delta(p,q,w)=\Delta(N)$ where $N(cd)=m_{cd}(v_cv_d)$ for $cd\ne12$ and, say, $N(12)=0$. 
The last value is immaterial because the derivative overwrites it. Then
\begin{equation}\label{eq:step}
\rho_i-\rho_{i-1}=\E_{(p,q,w)}\bigl[(W-Z)(pq)\,\Delta(p,q,w)\bigr].
\end{equation}

Fix $w\in\tup V{s-2}$. To realize the derivative by our tests, round the
anchor graph once and independently round the two role-dependent adjacency
vectors of each non-anchor vertex. Different pairs may share these random
bits; we only need the correct distribution for each individual pair.

Let $b$ be a random graph on $[3,s]$ with independent edges, $\Pr[b(cd)=1]=m_{cd}(w_cw_d)$. For each $u\notin\{w_c\}$ let $X_u,Y_u\in\{0,1\}^{[3,s]}$ have independent coordinates with $\Pr[X_u(c)=1]=m_{1c}(uw_c)$ and $\Pr[Y_u(c)=1]=m_{2c}(uw_c)$, all variables independent. Let $\chi(u)=(X_u,Y_u)$ for $u\notin\{w_c\}$, $\chi(w_c)=\bot$, and $A=A_{b,\chi}$, a random element of $\cA\subseteq\cT$. For distinct non-anchors $p,q$, the graph $F_t(b,X_p,Y_q)$ has independent edges with exactly the probabilities of $N^{[t]}$. Hence $\E f(F_t(b,X_p,Y_q))=\phi(N^{[t]})$, so $\E\,\partial_bf(X_p,Y_q)=\Delta(p,q,w)$ and
\[
\E A(pq)=\tfrac12\bigl(\Delta(p,q,w)+\Delta(q,p,w)\bigr),
\]
while $A(pq)=0$ if $p$ or $q$ is an anchor.

The color $\bot$ makes all anchor-incident pairs contribute zero. Our inner
product nevertheless averages over all pairs in $V$, whereas the hybrid step
averages only over pairs avoiding the anchors. We now account for this
normalization explicitly. Summing over unordered pairs, and using that
$(W-Z)(pq)$ is a function of the unordered pair,
\[
\E_A\ip{W-Z}A=\frac1{n(n-1)}\sum_{(p,q)}(W-Z)(pq)\,\Delta(p,q,w),
\]
the sum over ordered pairs of distinct non-anchors. There are $(n-s+2)(n-s+1)$ of them, and $(n)_s=(n)_{s-2}(n-s+2)(n-s+1)$. Averaging over uniform $w$ and comparing with \eqref{eq:step},
\[
\E\ip{W-Z}A=\alpha\,(\rho_i-\rho_{i-1}),
\qquad \alpha=\frac{(n-s+2)(n-s+1)}{n(n-1)}.
\]
Here $\alpha$ is the proportion of ordered pairs avoiding the anchors. Since
$n\ge4s$,
\[
\alpha\ge\Bigl(\frac{n-s+1}{n}\Bigr)^2>\frac12.
\]
Every realization of $A$ lies in $\cT$, so $|\ip{W-Z}A|\le\tn{W-Z}$, and therefore $|\rho_i-\rho_{i-1}|\le2\tn{W-Z}$. Summing over $i$ gives $|\rej(W)-\rej(Z)|\le2K\tn{W-Z}$. By Fact~\ref{fact:cost} and $\ones-2G\in\cT$, $|\cost_G(W)-\cost_G(Z)|=|\ip{W-Z}{\ones-2G}|\le\tn{W-Z}$. Add twice the first bound.
\end{proof}
\subsection{Boosting}

The counting lemma reduces the task to making every test correlation small.
We do this by repeatedly correcting a correlation that is still too large.
The sum of the corrections is allowed to be real-valued; clipping converts
it into valid edge probabilities. Keeping the un-clipped sum is useful
because it is described by a short list of tests.

Define $\clip:\Real\to[0,1]$ by $\clip(u)=\min\{1,\max\{0,u\}\}$,
and apply it pointwise to real-valued functions. To track progress, we use a
function whose derivative is exactly this clipping operation. Define
$\psi:\Real\to[0,\infty)$ by 
\[
\psi(u)=\begin{cases}0&u\le0,\\ u^2/2&0\le u\le1,\\ u-\tfrac12&u\ge1.\end{cases}
\]
By direct differentiation, $\psi\in C^1$, and $\psi'=\clip$ is $1$-Lipschitz. The following two properties explain this choice. The first will make our
potential nonnegative; the second bounds the cost of taking an update step: 
\begin{equation}\label{eq:psi}
\psi(u)=\sup_{0\le h\le1}\bigl(uh-\tfrac12h^2\bigr),\qquad \psi(u+r)\le\psi(u)+r\,\psi'(u)+\tfrac12r^2 .
\end{equation}
Indeed, the supremum is attained at $h=\clip(u)$, and the inequality follows from \[\psi(u+r)-\psi(u)-r\psi'(u)=\int_0^r(\psi'(u+t)-\psi'(u))\,dt.\]

For a target $H\in\cW_V$, define the potential
$\Phi_H:\Real^{\binom V2}\to[0,\infty)$, where $Q$ is the un-clipped
sum of the current corrections, by 
\[
\Phi_H(Q)=\E_e\Bigl[\psi\bigl(Q(e)\bigr)-Q(e)H(e)+\tfrac12H(e)^2\Bigr].
\]
It is nonnegative by \eqref{eq:psi} with $h=H(e)$.

When $Q$ already takes values in $[0,1]$, this potential is
$\tfrac12\lVert Q-H\rVert_2^2$. Outside that interval, it remains nonnegative
and its derivative with respect to $Q$ is $\clip(Q)-H$. This is the residual
we can detect using tests. Thus the potential continues to track useful
progress even when the un-clipped sum leaves the interval of valid weights.

The next lemma gives a short description of an approximation to any target
$H$: fewer than $\gamma^{-2}$ signed tests suffice. The number of vertices
and the number of available tests do not enter this bound.

\begin{lemma}\label{lem:boost}
Let $\cT$ be a finite nonempty family of functions $\binom V2\to[-1,1]$, let $G,H\in\cW_V$ and $0<\gamma<1$. There exist an integer $0\le t<\gamma^{-2}$ and $B_1,\dots,B_t\in\cT\cup(-\cT)$ such that
\(W:=\clip\Bigl(G+\gamma\sum_{j=1}^tB_j\Bigr)\in\cW_V\) satisfies \(\tn{H-W}\le\gamma\).
\end{lemma}
\begin{proof}
If a test sees an error larger than $\gamma$, move a distance $\gamma$ in the
sign of that test which reduces the error. The resulting decrease in the
potential is at least $\gamma^2/2$, so only a bounded number of such moves is
possible.

Set $Q_0=G$. Given $Q_j$, let $W_j=\clip(Q_j)$. If $\tn{H-W_j}\le\gamma$, stop. Otherwise some $A\in\cT$ has $|\ip{H-W_j}A|>\gamma$; choose $B_{j+1}\in\{\pm A\}$ with $\ip{H-W_j}{B_{j+1}}>\gamma$ and set $Q_{j+1}=Q_j+\gamma B_{j+1}$. Applying \eqref{eq:psi} pointwise, with $\psi'(Q_j)=W_j$ and $\lVert B_{j+1}\rVert_2\le1$,
\[
\Phi_H(Q_{j+1})\le\Phi_H(Q_j)-\gamma\ip{H-W_j}{B_{j+1}}+\tfrac{\gamma^2}2\lVert B_{j+1}\rVert_2^2<\Phi_H(Q_j)-\tfrac{\gamma^2}2 .
\]
Also $\Phi_H(Q_0)=\tfrac12\lVert G-H\rVert_2^2\le\tfrac12$ and
$\Phi_H\ge0$. After any positive number $t$ of updates,
\[
0\le\Phi_H(Q_t)<\tfrac12-\tfrac{t\gamma^2}2,
\]
so $t<\gamma^{-2}$. The process must therefore stop within this bound.
If no update is needed, $t=0$ already has the required properties.
\end{proof}

This is gradient descent for the convex potential $\Phi_H$, restricted to test directions. The step count $\gamma^{-2}$ is independent of $n$.

We now turn the short list of corrections into information stored at vertices.
In each step there is common information, i.e., the kind of test, its sign, and
possibly its anchor graph, and, for a derivative test, a color for each
vertex. We call the common information a \emph{symbol}. A vertex label stores
one symbol and one color for every possible step, padding unused steps by an
empty symbol. This is why the label length depends on the number of boosting
steps and on $s$, but not on $n$.

Here and below, fix $0<\gamma<1$, and use the family $\cT(G)$ constructed in
the preceding subsection.

Let $L=\lceil\gamma^{-2}\rceil$. The set of \emph{symbols} is
$\cC=\{\varnothing\}\ \sqcup\ \bigl(\{\mathrm{cost}\}\times\{\pm1\}\bigr)\ \sqcup\ \bigl(\{\mathrm{der}\}\times\{\pm1\}\times\cG_{[3,s]}\bigr)$,
and the \emph{label set} is $\Lambda=(\cC\times\Omega)^L$. A label $z\in\Lambda$ has $L$ components $(c_\ell(z),\omega_\ell(z))\in\cC\times\Omega$, and $\log|\Lambda|=L(\log|\cC|+\log|\Omega|)=O(s^2\gamma^{-2})$.

The decoder must recover an edge weight from the two endpoint labels and
the original edge. A genuine boosting description has matching common
symbols at every vertex. We nevertheless define the decoder on arbitrary
pairs of labels; the later CSP optimization must be allowed to choose any
label at any vertex. If the common symbols disagree, keeping the original
edge is a convenient way to return a valid weight.

Define the \emph{decoder} $D:\{0,1\}\times\Lambda\times\Lambda\to[0,1]$ as follows. For $g\in\{0,1\}$ and $z,z'\in\Lambda$:
\begin{itemize}
\item if $c_\ell(z)\ne c_\ell(z')$ for some $\ell$, then $D(g,z,z')=g$;
\item otherwise, let $c_\ell=c_\ell(z)=c_\ell(z')$ and set
\[
\mu_\ell=\begin{cases}0&c_\ell=\varnothing,\\ \sigma(1-2g)&c_\ell=(\mathrm{cost},\sigma),\\ \sigma\,\kappa_b\bigl(\omega_\ell(z),\omega_\ell(z')\bigr)&c_\ell=(\mathrm{der},\sigma,b),\end{cases}
\qquad
D(g,z,z')=\clip\Bigl(g+\gamma\sum_{\ell=1}^L\mu_\ell\Bigr).
\]
\end{itemize}
The decoder $D$ is symmetric in $(z,z')$. For a labeling $z:V\to\Lambda$,
write $z_v=z(v)$ and define $W_z\in\cW_V$ by
$W_z(uv)=D\bigl(G(uv),z_u,z_v\bigr)$.

In particular, every labeling defines a valid weighted graph, even if its
symbols are inconsistent. Consistency is needed only for encoding the
specific boosting approximation whose existence we have proved.

The next lemma connects the two descriptions: each boosting approximation
can be reconstructed from vertex labels. Its final assertion is what makes
sampling possible, since a sampled edge can be decoded without knowing any
labels outside its endpoints.

\begin{lemma}\label{lem:encode}
For every $G\in\cG_V$ and $H\in\cW_V$ there is $z\in\Lambda^V$ with $\tn{H-W_z}\le\gamma$, where $\cT=\cT(G)$. Moreover $W_z(uv)$ depends only on $G(uv)$, $z_u$, $z_v$.
\end{lemma}
\begin{proof}
Record the boosting sequence component by component, duplicating each step's
common symbol at every vertex and keeping its vertex-dependent color locally.
Apply Lemma~\ref{lem:boost} to get $B_j=\sigma_jA_j$, $j\le t\le L$. In component $j$, give every vertex the symbol $(\mathrm{cost},\sigma_j)$ and color $\bot$
if $A_j=\ones-2G$. Otherwise $A_j=A_{b_j,\chi_j}$; give every vertex the symbol $(\mathrm{der},\sigma_j,b_j)$ and give vertex $v$ the color $\chi_j(v)$. Components $j>t$ get $(\varnothing,\bot)$. All symbols agree across vertices, so $D$ is in its second case with $\mu_j=B_j(uv)$ for every $uv$. Thus $W_z=\clip(G+\gamma\sum_jB_j)$, the function of Lemma~\ref{lem:boost}. The last claim is the definition of $D$.
\end{proof}

We can now restrict the search for a repair to weighted graphs specified by
labels. Define
\[
\Mt(G)=\min_{z\in\Lambda^V}\cE_G(W_z).
\]
The next proposition says that this restriction loses little: encode a true
minimizer of $M(G)$ and use the counting lemma to bound the change in its
objective value. Arbitrary labelings cannot artificially improve on $M(G)$,
because they still decode to genuine weighted graphs.

\begin{proposition}\label{prop:finite}
For $n\ge4s$,  \(M(G)\le\Mt(G)\le M(G)+(1+4K)\gamma,\) and if $\cP_V\ne\emptyset$,
\[
\dist(G,\cP)-\tau\ \le\ \Mt(G)\ \le\ \dist(G,\cP)+2\tau+(1+4K)\gamma .
\]
\end{proposition}
\begin{proof}
The lower bound is immediate from inclusion of the feasible sets; the upper
bound follows by approximating one minimizing weighted graph.
Since $W_z\in\cW_V$, we have $\Mt(G)\ge M(G)$. Let $H^*$ attain $M(G)$ and take $z$ from Lemma~\ref{lem:encode} for $H=H^*$. Lemma~\ref{lem:counting} gives $\Mt(G)\le\cE_G(W_z)\le M(G)+(1+4K)\gamma$. The distance bounds follow from Lemma~\ref{lem:penalty}.
\end{proof}

Observe that the boosting sequence is never computed by the algorithm. It only proves that a short-description repair of nearly optimal energy exists.

\subsection{Estimating the optimum from a sample}

The labels have reduced our search to a finite optimization problem. To apply
a sampling theorem, we must express its objective as an average of bounded
functions of a few vertex labels. The tester term already depends on $s$
vertices. We write the edit term on the same tuples by averaging its pair
costs over the $K$ pairs in each tuple.

A \emph{dense $s$-ary constraint satisfaction problem (CSP)} consists of variables $V$, a label set $\Lambda$, and for each $\mathbf v\in\tup Vs$ a predicate $P_{\mathbf v}:\Lambda^s\to[0,1]$. For $U\subseteq V$ with $|U|=m$ set $\Val(U)=\max_{z\in\Lambda^U}\ m^{-s}\sum_{\mathbf v\in\tup Us}P_{\mathbf v}(z_{v_1},\dots,z_{v_s})$, and let $\OPT=\Val(V)$. Here a predicate is a score in $[0,1]$, so the goal
is to maximize the average score. We take $m\ge s$ whenever using this value.
The normalization is $m^{-s}$, rather than $1/(m)_s$: equivalently, tuples
with a repeated vertex have score zero. This convention is responsible for
the factor $a_n$ below.

For our graph $G$, a tuple score will be one minus a rescaled local loss.
The division by $3$ is enough because the edit loss is at most $1$ and the
rejection penalty at most $2$. For $\mathbf v\in\tup Vs$ and $\zeta=(\zeta_1,\dots,\zeta_s)\in\Lambda^s$ let $Y\in\cW_s$ be $Y(ab)=D\bigl(G(v_av_b),\zeta_a,\zeta_b\bigr)$. Define the loss $\ell_{\mathbf v}:\Lambda^s\to[0,3]$ and the predicate $P_{\mathbf v}$ by $P_{\mathbf v}:=1-\tfrac13\ell_{\mathbf v}$, with
\[
\ell_{\mathbf v}(\zeta)=\frac1K\sum_{ab\in\binom{[s]}2}\Bigl[G(v_av_b)\bigl(1-Y(ab)\bigr)+\bigl(1-G(v_av_b)\bigr)Y(ab)\Bigr]+2\,\phi(Y).
\]
The loss depends only on $G[\mathbf v]$ and $\zeta$. Denote this CSP by
$\mathcal I(G)$. In $\Val(U)$ we always restrict this same instance to $U$:
the alphabet $\Lambda$, decoder $D$, and rejection rule $f_n$ are kept fixed.
This restriction need not be the instance one would construct by treating
$G[U]$ as a new input graph.

Let $a_n=(n)_s/n^s$, the probability that $s$ independent uniformly chosen
vertices of $V$ are all distinct. The next lemma verifies that the local
losses average to the objective we want, with precisely this normalization
factor when losses are converted to CSP scores.

\begin{lemma}\label{lem:csp}
For every $z\in\Lambda^V$, $\E_{\mathbf v}\,\ell_{\mathbf v}(z_{v_1},\dots,z_{v_s})=\cE_G(W_z)$, with $\mathbf v$ uniform in $\tup Vs$. Hence
\[
\OPT(\mathcal I(G))=a_n\bigl(1-\Mt(G)/3\bigr).
\]
\end{lemma}
\begin{proof}
A fixed pair of positions in a uniform injective tuple is a uniform pair of
vertices. Consequently, distributing the edit loss over larger tuples does
not change its mean; the rejection term agrees by definition.
For uniform $\mathbf v$ and fixed $ab$, $v_av_b$ is a uniform pair, so the $ab$-term of the first part has expectation $\E_e[G(e)(1-W_z(e))+(1-G(e))W_z(e)]=\cost_G(W_z)$ (Fact~\ref{fact:cost}). Averaging over $ab$ gives $\cost_G(W_z)$. Since $Y=W_z[\mathbf v]$ here, the second part has expectation $2\rej(W_z)$. For $\OPT$: $m=n$ and $|\tup Vs|=a_nn^s$, so $n^{-s}\sum_{\mathbf v}P_{\mathbf v}=a_n(1-\cE_G(W_z)/3)$; maximize over $z$.
\end{proof}

We use the following external sampling theorem. It controls the optimum
\emph{after} optimizing the labels on the sample, not just the sampled value
of a fixed labeling. This distinction matters: the best labeling of a sample
can depend on the sample and need not extend to an equally good labeling of
the whole graph. The guarantee is in expectation; independent repetition
will turn it into the desired probability guarantee.

\begin{theorem}[{Bernasconi--Castiglioni--Celli--Farina--Malavolta~\cite[Theorem~3.3]{BCCFM}}]\label{thm:sub}
There are absolute constants $c_*,C_*>0$ such that for every dense $k$-ary CSP with label set of size $Q_0$ and every $0<\theta<1$, if
\begin{align*}
m&=\bigl\lceil c_*\,k^{12}\theta^{-6}\log^2(Q_0k^2/\theta)\bigr\rceil,\\
n&\ge c_*^2k^{24}\theta^{-13}\log^4(Q_0k^2/\theta),
\end{align*}
$n\ge m$, and $U\subseteq V$ is a uniformly random $m$-subset, then $\bigl|\E\Val(U)-\OPT\bigr|\le C_*\theta$.
\end{theorem}

The dependence on the label set is only through $\log Q_0$. This is essential, since $|\Lambda|$ is exponential in $\operatorname{poly}(s,1/\varepsilon)$.

The penalty and label approximation account for a small deterministic
error; the sampled CSP optima account for the remaining random error.
The threshold $N_0$ ensures that all counting and sampling hypotheses hold.
Graphs below that threshold can be read in full at a polynomial query cost.

\begin{proof}[Proof of Theorem~\ref{thm:main}]

Let $\varepsilon\in(0,1)$, put $\tau=\varepsilon/100$, and choose $s$ as in
Proposition~\ref{prop:normal}, so that
\[
s=O\bigl(\max\{1,q(\varepsilon/100)\}\log(100/\varepsilon)\bigr).
\]
Set $K=\binom s2$ and $\gamma_0=\varepsilon/[100(1+4K)]$.
Choose a rational $\gamma\in[\tfrac12\gamma_0,\gamma_0]$.
Then $(1+4K)\gamma\le\varepsilon/100$ and
\begin{equation}\label{eq:logq}
\log|\Lambda|=O(s^2\gamma^{-2})=O(s^6\varepsilon^{-2}).
\end{equation}
For $n\ge4s$ and $\cP_V\ne\emptyset$, Proposition~\ref{prop:finite} gives
\begin{equation}\label{eq:F}
|\Mt(G)-\dist(G,\cP)|\le\tfrac{3\varepsilon}{100}.
\end{equation}
Apply Theorem~\ref{thm:sub} with $k=s$, $Q_0=|\Lambda|$ and $\theta=c_0\varepsilon$, where $c_0$ is an absolute constant with $\theta<1$ and $C_*\theta\le\varepsilon/24$. Let $m$ be the resulting sample size and $N_0$ the maximum of the theorem's threshold on $n$, of $m$, of $4s$ and of $2s^2$ (rounded up). By \eqref{eq:logq}, both $m$ and $N_0$ are $\operatorname{poly}(s,1/\varepsilon)$ with absolute degrees. Let $T=\lceil432\varepsilon^{-2}\rceil$ be the number of independent
samples.

We may identify the input vertex set with $[n]$. The rational choice of
$\gamma$, together with the rational values of $f_n$, makes the finite
sample objectives rational, so their optima can be computed by exhaustive
search without an additional approximation error.

\paragraph{The algorithm}
\begin{enumerate}[label=\arabic*.]
\item By finite enumeration decide whether $\cP_{[n]}=\emptyset$; if so output $1$. If $n<2$ output $0$.
\item If $2\le n\le N_0$: query all pairs and output $\dist(G,\cP)$ exactly.
\item If $n>N_0$: draw independent uniform $m$-subsets $U_1,\dots,U_T\subseteq[n]$ and query all pairs inside each. By exhaustive search compute $\Val_i=\Val(U_i)\in[0,1]$ for the CSP $\mathcal I(G)$ restricted to $U_i$ (each predicate depends only on $G[\mathbf v]$, the function $f_n$ and $D$). With $\overline{\Val}=T^{-1}\sum_{i=1}^T\Val_i$, output $\widehat d=\clip\bigl(3(1-\overline{\Val}/a_n)\bigr)$.
\end{enumerate}

\paragraph{Correctness for $n>N_0$}
The sampling theorem bounds the bias of each sampled optimum. Averaging
independent copies controls its fluctuation around that mean. We then invert
the exact CSP identity from Lemma~\ref{lem:csp}; the lower bound on $a_n$
ensures that this inversion enlarges the error by only a constant factor.

By Theorem~\ref{thm:sub} and Lemma~\ref{lem:csp}, $|\E\Val_i-\OPT(\mathcal I(G))|\le\varepsilon/24$. The $\Val_i$ are independent with values in $[0,1]$, so $\operatorname{Var}(\Val_i)\le1/4$ and
$\operatorname{Var}(\overline{\Val})\le1/(4T)$. Chebyshev's inequality gives
\[
\Pr\bigl[|\overline{\Val}-\E\Val_i|>\varepsilon/24\bigr]\le\frac{144}{T\varepsilon^2}\le\frac13 .
\]
So with probability at least $2/3$, $|\overline{\Val}-\OPT(\mathcal I(G))|\le\varepsilon/12$. Since $n>2s^2$, $a_n=\prod_{j<s}(1-j/n)\ge1-\binom s2/n\ge\tfrac12$. By Lemma~\ref{lem:csp}, $\Mt(G)=3(1-\OPT(\mathcal I(G))/a_n)$, so the quantity inside the clip differs from $\Mt(G)$ by at most $3\cdot\frac\varepsilon{12}\cdot2=\frac\varepsilon2$. With \eqref{eq:F} the total error is below $\frac\varepsilon2+\frac{3\varepsilon}{100}<\varepsilon$, and clipping to $[0,1]\ni\dist(G,\cP)$ cannot increase it. For $n\le N_0$ the output is exact.

\paragraph{Query complexity}
All graph queries are made inside the sampled vertex sets, or inside the
whole graph when its order is at most $N_0$. Searching through labels and
evaluating their predicates uses no further graph queries. Thus the number
of queries is at most
\[
\max\Bigl\{\tbinom{N_0}2,\ T\tbinom m2\Bigr\}=\operatorname{poly}(s,1/\varepsilon),
\]
and substituting $s=O(\max\{1,q(\varepsilon/100)\}\log(100/\varepsilon))$ proves the theorem with $c=1/100$.
\end{proof}

For the corollary, the point is that a natural property already requires
many queries to test at small proximity. Thus the explicit accuracy parameter
in the estimator's bound can be paid for by the tester's own query complexity.
We use the definitions of natural and non-trivial hereditary properties from
Section~\ref{sec:introduction}.

\begin{proof}[Proof of Corollary~\ref{cor:main}]
For natural properties, \cite[Proposition~1.10]{GKS} gives the edge-query
lower bound $q(\eta)=\Omega(1/\eta)$. For non-trivial hereditary properties,
\cite[Proposition~1.7]{GKS} gives this lower bound for the canonical vertex
sample size. The Goldreich--Trevisan conversion~\cite{GT} turns a tester with
$q(\eta)$ edge queries into one with $O(q(\eta))$ sampled vertices, so the
same lower bound follows for our edge-query notation.
Taking $\eta=c\varepsilon$, we obtain
$1/\varepsilon=O(q(c\varepsilon))$ for sufficiently small $\varepsilon$.
This also absorbs the $\max\{1,\cdot\}$ in Theorem~\ref{thm:main}, proving
the corollary.
\end{proof}
\subsection*{AI disclosure}
AI-assisted tools, specifically \emph{Quillbot}, were utilized only to enhance the exposition's clarity and language. The mathematical arguments were not produced using them. Additionally, the author made use of the optional AI feedback for STOC 2027. The system found a few typos and some formatting errors, which the author personally corrected.

\subsection*{Acknowledgments}
This project started while the authors were participating in the 2025 NII International Internship Program.  The authors are grateful to the program for its support.  G.C. is supported by the CB European PhD Studentship funded by Trinity College, Cambridge, and would like to thank the Isaac Newton Institute for Mathematical Sciences, Cambridge, for support and hospitality during the programme Logical Foundations of Complexity Theory, where part of the work on this paper was undertaken. This work was supported by EPSRC grant EP/Z000580/1.


\begin{thebibliography}{99}

\bibitem{Goldreich}
O.~Goldreich,
\emph{Introduction to Property Testing},
Cambridge University Press, 2017.

\bibitem{GT}
O.~Goldreich and L.~Trevisan,
Three theorems regarding testing graph properties,
\emph{Random Structures \& Algorithms} 23 (2003), 23--57.

\bibitem{GGR}
O.~Goldreich, S.~Goldwasser and D.~Ron,
Property testing and its connection to learning and approximation,
\emph{Journal of the ACM} 45 (1998), 653--750.

\bibitem{AS}
N.~Alon and A.~Shapira,
A characterization of the (natural) graph properties testable with one-sided error,
\emph{SIAM Journal on Computing} 37 (2008), 1703--1727.

\bibitem{Gowers}
W.~T.~Gowers,
Lower bounds of tower type for Szemer\'edi's uniformity lemma,
\emph{Geometric and Functional Analysis} 7 (1997), 322--337.

\bibitem{RD}
V.~R\"odl and R.~Duke,
On graphs with small subgraphs of large chromatic number,
\emph{Graphs and Combinatorics} 1 (1985), 91--96.

\bibitem{CSY}
G.~Carenini, C.~Seth and Y.~Yoshida,
A quantitative container characterization of one-sided testability,
2026, \href{https://arxiv.org/abs/2608.01523}{arXiv:2608.01523}.

\bibitem{BS}
E. Blais and C.~Seth,
Testing graph properties with the container method, 
\emph{SIAM Journal on Computing} 54 6 (2025), 1626-1642.

\bibitem{KS}
N. Kushnir and A. Shapira, 
Towards tight bounds for testing $k$-colorability, 
\emph{Proceedings of APPROX/RANDOM} (2026), 35:1--35:13. 


\bibitem{BMS}
J.~Balogh, R.~Morris and W.~Samotij,
Independent sets in hypergraphs,
\emph{Journal of the American Mathematical Society} 28 (2015), 669--709.

\bibitem{ST}
D.~Saxton and A.~Thomason,
Hypergraph containers,
\emph{Inventiones Mathematicae} 201 (2015), 925--992.

\bibitem{PRR}
M.~Parnas, D.~Ron and R.~Rubinfeld,
Tolerant property testing and distance approximation,
\emph{Journal of Computer and System Sciences} 72 (2006), 1012--1042.

\bibitem{FF}
E.~Fischer and L.~Fortnow,
Tolerant versus intolerant testing for Boolean properties,
\emph{Theory of Computing} 2 (2006), 173--183.

\bibitem{FN}
E.~Fischer and I.~Newman,
Testing versus estimation of graph properties,
\emph{SIAM Journal on Computing} 37 (2007), 482--501.

\bibitem{HKLFS}
C.~Hoppen, Y.~Kohayakawa, R.~Lang, H.~Lefmann and H.~Stagni,
On the query complexity of estimating the distance to hereditary graph properties,
\emph{SIAM Journal on Discrete Mathematics} 35 (2021), 1238--1251.

\bibitem{FR}
N.~Fiat and D.~Ron,
On efficient distance approximation for graph properties,
\emph{Proceedings of SODA 2021}, 1618--1637.

\bibitem{GKS}
L.~Gishboliner, N.~Kushnir and A.~Shapira,
Testing versus estimation of graph properties, revisited,
\emph{Random Structures \& Algorithms} 65 (2024), 460--487.

\bibitem{FK}
A.~Frieze and R.~Kannan,
Quick approximation to matrices and applications,
\emph{Combinatorica} 19 (1999), 175--220.

\bibitem{DT}
C.~Dwork and P.~Tankala,
Efficient and private property testing via indistinguishability,
\href{https://arxiv.org/abs/2511.03653v2}{arXiv:2511.03653v2}, 2026.

\bibitem{TTV}
L.~Trevisan, M.~Tulsiani and S.~Vadhan,
Regularity, boosting, and efficiently simulating every high-entropy distribution,
\emph{Proceedings of IEEE CCC}, 2009.

\bibitem{BCCFM}
M.~Bernasconi, M.~Castiglioni, A.~Celli, G.~Farina and G.~Malavolta,
A subsampling theorem for constraint satisfaction problems with large arity,
2026, \href{https://arxiv.org/abs/2608.00203}{arXiv:2608.00203}.

\end{thebibliography}
\end{document}